\documentclass{ws-p8-50x6-00}
\usepackage{epsfig}
\begin{document}

\title{Can we determine the nuclear equation of state from 
heavy ion collisions?}

\author{T. Gaitanos, H.H. Wolter}
\address{Sektion Physik, Universit\"at M\"unchen, 
D-85748 Garching, Germany}
% \\E-mail: Theo.Gaitanos@Physik.uni-muenchen.de}
\author{C. Fuchs, A. Faessler}
\address{Institut f\"ur Theor. Physik, Universit\"at T\"ubingen, 
D-72076 T\"ubingen, Germany}

%%%%%%%%%%%%%%%%%%%%%%%%%%%%%%%%%%%%%%%%%%%%%%%%%%%%%%%%%%%%%%
% You may repeat \author \address as often as necessary      %
%%%%%%%%%%%%%%%%%%%%%%%%%%%%%%%%%%%%%%%%%%%%%%%%%%%%%%%%%%%%%%

\maketitle

\abstracts{We discuss the problems involved in extracting the nuclear 
equation-of-state from heavy-ion collisions. We demonstrate that the 
equation of state becomes effectively softer in non-equilibrium and this 
effect is observable in terms of collective flow effects. Thus, non-equilibrium 
effects must be included in transport descriptions on the level of the effective 
mean fields. A comparison with transverse momentum, rapidity, and centrality 
selected flow data show the reliability and limitations of the underlying 
interaction which was derived from microscopic Dirac-Brueckner (DB) results.}
%%%%%%%%%%%%%%%%%%%%%%%%%%%%%%%%%%%%%%%%%%%%%%%%%%%%%%%%%%%%%%%%%%%%%%%%
%%%%%%%%%%%%%%%% TEXT
%%%%%%%%%%%%%%%%%%%%%%%%%%%%%%%%%%%%%%%%%%%%%%%%%%%%%%%%%%%%%%%%%%%%%%%%

Heavy-ion collisions open the possibility to explore the nuclear 
equation-of-state (EOS) far away from saturation. In particular, the 
collective flow is  connected to the 
dynamics during the high density phase of such reactions and thus yields 
information on the nuclear EOS \cite{stoecker86}. The different components of 
collective 
flow, in particular, the in-plane and out-of-plane flow \cite{andronic99} has recently 
attracted great interest 
\cite{andronic99,dani00,pinkenburg99}, since also the 
energy, centrality and transverse momentum dependence has been measured. 

From the theoretical point of view, however, the determination of the EOS, 
which is an equilibrium concept, is not straightforward in heavy ion 
collisions, which are highly non-equilibrium processes. An appropriate starting 
point are the Kardanoff-Baym equations \cite{Kard} for the non-equilibrium 
real-time Green functions. To arrive at transport descriptions several 
approximations are usually introduced: the semi-classical approximation 
neglecting non-locality and memory effects, the quasi-particle approximation 
neglecting the finite width of the spectral functions, and the T-Matrix 
approximation, specifying the self-energies in the Brueckner ladder 
approximation. Several attemps have recently been made to improve the first 
two approximations \cite{cei}. They are expected to be important 
e.g. for subthreshold particle production, but it is not known whether they 
are also relevant for flow observables.

With the above approximations one arrives at a BUU-type transport equation 
with DB self-energies which should be specified for the general non-equilibrium 
configuration. Since these cannot be calculated one further introduces a 
local nuclear matter approximation neglecting spatial variations and adopting 
a model for the local nuclear matter phase space distribution. Two models have 
mainly been used: One is the Local Density Approximation (LDA) assuming 
equilibrated nuclear matter at the local total density, usually supplemented 
with an empirical momentum dependence taken from the real part of the optical 
model. The other is the Local phase space Configuration Approximation (LCA) 
which assumes a configuration of two interpenetrating streams of nuclear 
matter with a given relative velocity (also called colliding nuclear matter, 
CNM), i.e. a two-Fermi-sphere configuration with Pauli-effects taken into 
account. In the first case this is directly the nuclear EOS. For the second 
case the self energies have not been calculated in the DB approach (but 
for the non-relativistic case \cite{mueth}). However, we have 
developed a self-consistent extrapolation starting from nuclear matter 
DB calculations as discussed in detail in ref. \cite{sehn96}. This then 
gives the model non-equilibrium self energies as a function of the 
configuration parameters, namely the two Fermi momenta and the relative 
velocity. 
%%%%%%%%%%%%%%%%
\begin{figure}[t]
\begin{center}
\unitlength1cm
\begin{picture}(12,4.)
%\put(0,0){\framebox(12,4.){}}
\put(0.0,-0.5){\makebox{\psfig{file=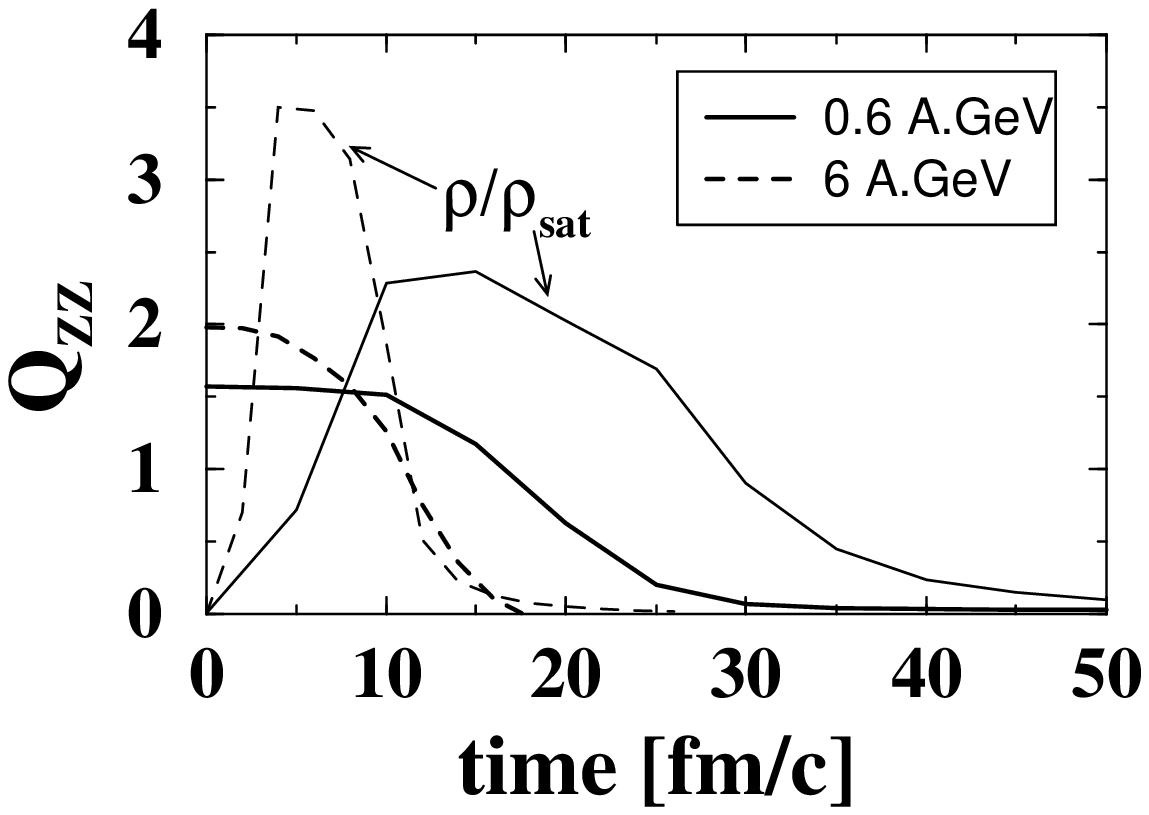,width=6.0cm}}}
\put(6,-0.5){\makebox{\psfig{file=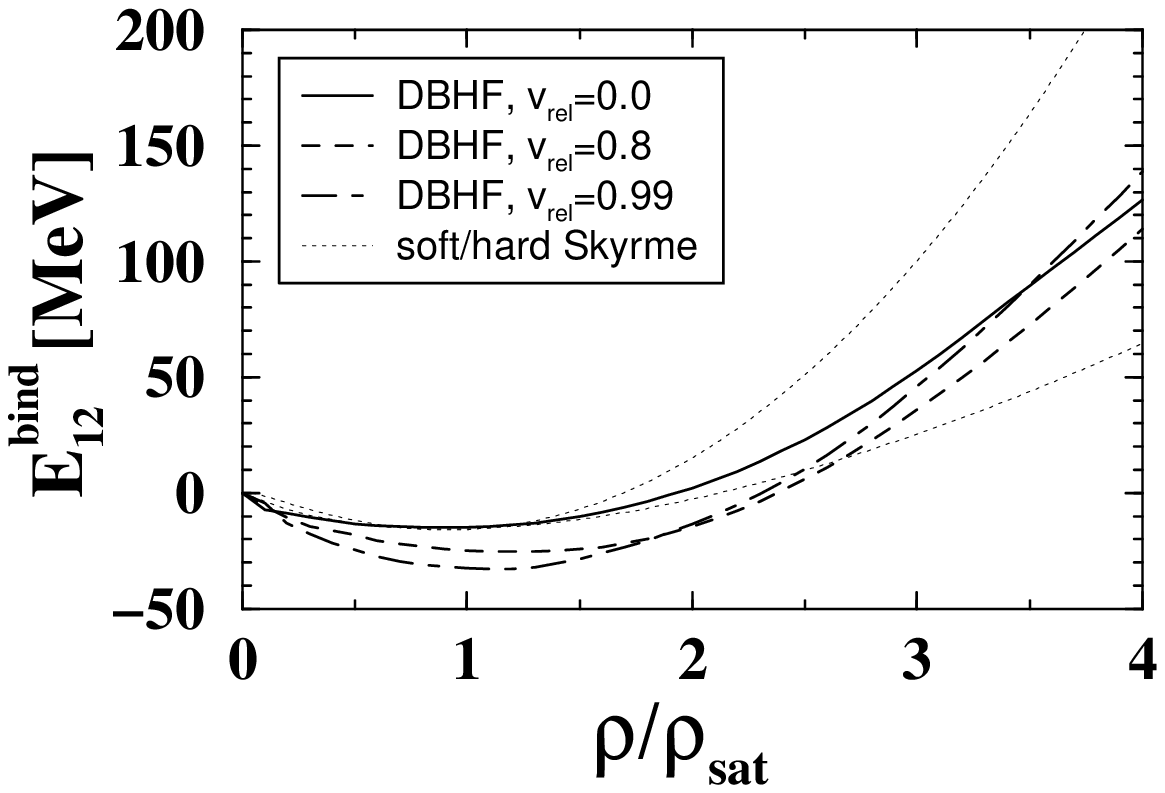,width=6.0cm}}}
\end{picture}
\caption{\label{fig1}
Left: Time evolution of the quadrupole moment of the energy-momentum tensor 
$Q_{zz}$ (bold lines) and of the 
central density (in units of $\rho_{\rm sat}$, dashed lines) at the center 
in central $Au+Au$ collisions at 0.6 and 6.0 A.GeV.
Right: Energy per particle of a two-Fermi-sphere configuration (CNM) (minus 
the relative kinetc energy) as a function of total density. This ``reduced'' 
EOS is shown for DB self energies for several relative velocities and for 
two (equilibrium) Skyrme parametrizations.}
\end{center}
\end{figure}
%%%%%%%%%%%%%%%%

The relevance of such non-equilibrium effects in the dynamical situation of heavy 
ion collisions 
is demonstrated in Fig.~\ref{fig1} (left side). Here the quadrupole moment 
of the  energy-momentum tensor 
$Q_{zz} = \frac{2T^{33} - T^{11} -T^{22}}{T^{33} + T^{11} + T^{22}}$
is shown together with the baryon density as a function of time. It is clearly 
seen that the local phase space 
is significantly anisotropic during times which are comparable with the 
compression phase 
of the collision independent of beam energy. Hence, non-equilibrium effects 
are present during a large part of the compression phase where one wants 
to study the EOS at 
high densities. Therefore non-equilibrium effects should be considered at 
least on the level of the effective in-medium interaction, i.e. the mean 
fields in transport calculations should be taken for non-equilibrium 
configurations instead for equilibrated nuclear matter. 
Actual transport calculations have verified that the anisotropic phase space in 
heavy ion reactions can be parametrized reliably by CNM configurations 
\cite{gait99}. 

In order to have a qualitative feeling for the non-equilibrium effects we 
define an effective, ``reduced'' EOS for the CNM configuration. This is 
given by subtracting from the binding energy per particle for the CNM 
configuration with a given relative velocity $v_{rel}$ the irrelevant 
kinetic energy of relative motion between the two streams. This 
quantity is shown on the right side of Fig.~\ref{fig1} for symmetric 
 CNM configurations ($k_{F_{1}}=k_{F_{2}}$) for relative velocities 
$v_{rel}=0.8(0.99)$ corresponding to 
beam energies of $0.6(6.0)$ GeV per nucleon as a function of the total 
density. $v_{rel}=0$ corresponds to the isotropic case 
(equilibrated 
nuclear matter, i.e. one Fermisphere). 
 As a reference we also show
the widely used Skyrme parameterisations with a 
soft/hard EOS with K = 200/380 MeV \cite{stoecker86}. 

It is seen that the effective EOS becomes more 
attractive, i.e. softer, compared to the equilibrium EOS and 
saturates at a higher 
total density. This effect is understood qualitativelyin the following way: 
If there was no interaction between the two nuclear matter streams 
then the reduced energy per particle of the two-Fermi-sphere 
configuration were just the energy per particle of one Fermi 
sphere at half the total denisity. The minimum of the reduced 
EOS would then be shifted to twice $\rho_{sat}$. The interaction 
energy per particle can be estimated as twice the real part of the 
optical potential at the relative momentum which can be estimated 
as twice the Fermi momentum at the corresponding density. For 
saturation density this is attractive but decreases with 
increasing momentum. Thus we expect a minimum of the reduced 
EOS between one and twice $\rho_{sat}$ and below the saturation 
energy of equilibrium nuclear matter. This is what is seen 
qualitatively in Fig. 1.
%%%%%%%%%%%%%%%%
\begin{figure}[t]
\begin{center}
\unitlength1cm
\begin{picture}(12,7.3)
%\put(0,0){\framebox(12,7.3){}}
\put(0.5,-0.5){\makebox{\psfig{file=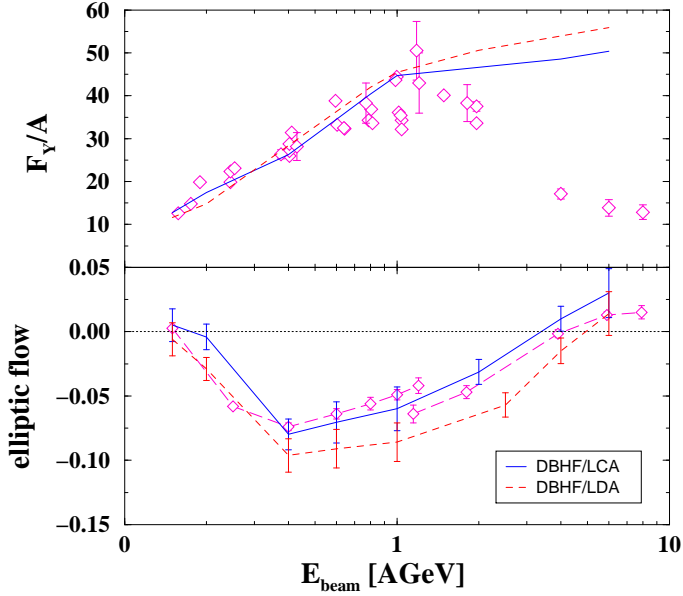,width=9.0cm}}}
\end{picture}
\caption{\label{fig2} 
Flow per particle $F_y/A$ and 
elliptic flow $v_{2}$ at mid-rapidity ($|\Delta Y^{(0)}|<0.1$) in 
semi-central $Au+Au$ reactions from SIS to AGS energies. The transport 
calculations \protect\cite{fuchs00} with microscopic DB mean fields in the LCA (solid) and 
LDA (dashed) approaches are compared to the data (symbols) 
\protect\cite{andronic99,pinkenburg99}.}
\end{center}
\end{figure}
%%%%%%%%%%%%%%%%

The question of interest is now whether these non-equilibrium 
effects are observable in terms 
of 
collective flow. We have therefore performed transport 
calculations 
using CNM self energies derived from DB mean fields calculated 
by the T\"ubingen-group (DBF) \cite{dbhf} in the two approximations: 
the LDA which refers to equilibrated nuclear matter (DBF/LDA), and the 
local configuration approximation (LCA) where the non-equilibrium 
DBF mean fields are used in the transport calculations (DBF/LCA). 

We discuss azimuthal distributions as function of transverse momentum, rapidity, 
centrality and beam energy. Such distributions have been parametrized in terms of 
a Fourier series $dN/d\phi=a_{0}(1+2a_{1}cos(\phi)+2a_{2}cos(2\phi))$ with $\phi$ 
the azimuthal angle with respect to the reaction plane. The coefficients 
$a_{1}$ and $a_{2}$ describe
the in-plane and out-of-plane components of collective flow, respectively, and they 
depend on transverse momentum $P_{t}^{(0)}=(P/A)/(P_{cm}^{proj}/A_{proj})$, 
rapidity $Y^{(0)}=Y_{cm}/Y_{cm}^{proj}$, centrality and energy. 
The in-plane or directed flow $v_{1}$ can also be characterized by the quantity 
$F_{y}=\frac{d<k_{x}(Y^{(0)})>/A}{dY^{(0)}} \bigg|_{Y^{(0)}=0}$ as the slope of 
the mean in-plane transverse momentum at mid rapidity. 
$v_{2}$ is often referred to
as the elliptic flow. Its sign describes the transition from out-of-plane ($v_{2}<0$) 
to in-plane flow ($v_{2}>0$) \cite{pinkenburg99}. The squeeze-out ratio $R_{N}$ 
is connected to the elliptic flow through $R_{N}=(1-2v_{2})/(1+2v_{2})$. 
%%%%%%%%%%%%%%%%
\begin{figure}[t]
\begin{center}
\unitlength1cm
\begin{picture}(12,4.75)
%\put(0,0){\framebox(12,4.75){}}
\put(0.5,-0.5){\makebox{\psfig{file=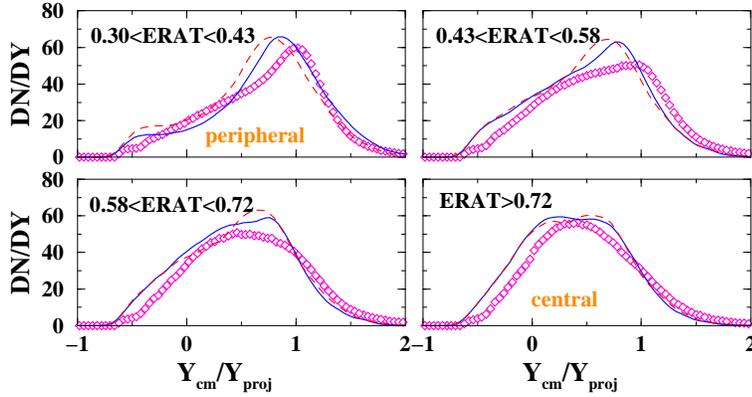,width=10.0cm}}}
\end{picture}
\caption{\label{fig3}
Rapidity distributions for $Au+Au$ collisions at 400 AMeV selected 
according to bins of ERAT which is a measure of centrality (increasing 
with increasing centrality). Calculations are shown for the LCA (solid) 
and LDA (dashed) approximations for the self energies and compared with 
the data (symbols) \protect\cite{fopi95}.}
\end{center}
\end{figure}
%%%%%%%%%%%%%%%%
Fig.~\ref{fig2} shows the observables $F_y/A$ and $v_{2}$ for $Au+Au$ collisions 
from SIS to AGS energies. Rather complete data are available from the EOS and 
FOPI collaborations, which are, however, not completely consistent for $v_2$. 
It is seen that the calculations \cite{fuchs00} reproduce the flow $F_{y}$ only for 
SIS energies, while at AGS energies the calculations overstimate the data.
This is to be expected because the DB self energies do not reproduce the 
empirical saturation of the real part of the optical potential at energies 
above about 1 GeV, but rather lead to linearly increasing repulsion. This
situation also appears in non-relativistic 
descriptions 
using momentum dependent Skyrme-parametrizations (see \cite{dani98}).
The difference between the LDA and LCA descriptions is not significant 
because the DBF self energies are not strongly momentum dependent at saturation 
density.

The description of the elliptic flow is generally much better because 
as a ratio it is not so strongly dependent on the absolute transverse flow.  
We see that both models generally
reproduce 
the qualitative trend of the data, i.e. the transition from small $v_2$ at 
0.2 A.GeV to 
a preferential out-of plane flow ($v_2 < 0$) which is maximal at 0.4 A.GeV with 
the 
subsequent transition back to in-plane flow around 4 A.GeV. There is now, 
however, a pronounced difference between the two approximations. The 
calculation without 
non-equilibrium effects (LDA) shows a larger squeeze-out effect and 
also has a larger transition 
energy from a out-of-plane to an in-plane emission of participant matter. This 
effect is reduced for the LCA calculation which
can be understood by the effective softening of the equation of state during the 
initial 
non-equilibrium phase of the collision as discussed in connection with
Fig.~\ref{fig1}. We also mention that the observed difference is of the 
same magnitude 
as the difference between a soft and hard EOS (see Fig.~\ref{fig2} of Ref. 
\cite{dani98}).
Furthermore the inclusion of the non-equilibrium effects considerably improves 
the agreement 
with the data which means that the underlying effective interaction (DB) is able 
to describe the dynamics of heavy ion collisions. 

In Fig.~\ref{fig3} we show rapidity distributions for $Au+Au$ collisions at 
400 AMeV selected according to a quantity ERAT, the ratio between transverse 
and longitudinal total momentum, which has been shown to be a measure of 
centrality \cite{fopi95}. The rapidity distributions are reasonably well 
reproduced, in particular for central collisions (large ERAT). A general 
tendency is that the spectator particles observed at $Y^{(0)}\simeq 1$ 
in the experiment are more stopped in the calculations. This effect is 
larger in LDA than LCA, because the field is more repulsive in the former 
case (see Fig. 1).

To summarise the effective equation-of-state probed by the compression phase in energetic 
heavy ion 
reactions is in a significant way governed by local non-equilibrium. The 
anisotropy 
of the phase space lowers the binding energy per particle and makes the effective 
EOS seen in 
heavy ion reactions softer. This fact is reflected in the behaviour of 
collective flow 
phenomena. 
We conclude that ``geometric'' phase space effects should be taken into account on 
the level 
of the effective interaction when conclusions on the equilibrium EOS are drawn 
from 
heavy ion collisions. The comparison with experiments shows that microscopic many-body 
methods methods that are 
successfull for nuclear structure also explain heavy ion reactions quantitatively 
at 
SIS energies \cite{fuchs96}.
\vspace*{-0.5cm}
%\section*{References}

\end{document}